# Meeting Effectiveness and Inclusiveness in Remote Collaboration

ROSS CUTLER, Microsoft Corp., USA
YASAMAN HOSSEINKASHI, Microsoft Corp., USA
JAMIE POOL, Microsoft Corp., USA
SENJA FILIPI, Microsoft Corp., USA
ROBERT AICHNER, Microsoft Corp., USA
YUAN TU, Microsoft Corp., USA
JOHANNES GEHRKE, Microsoft Corp., USA

A primary goal of remote collaboration tools is to provide effective and inclusive meetings for all participants. To study meeting effectiveness and meeting inclusiveness, we first conducted a large-scale email survey (N=4,425; after filtering N=3,290) at a large technology company (pre-COVID-19); using this data we derived a multivariate model of meeting effectiveness and show how it correlates with meeting inclusiveness, participation, and feeling comfortable to contribute. We believe this is the first such model of meeting effectiveness and inclusiveness. The large size of the data provided the opportunity to analyze correlations that are specific to sub-populations such as the impact of video. The model shows the following factors are correlated with inclusiveness, effectiveness, participation, and feeling comfortable to contribute in meetings: sending a pre-meeting communication, sending a post-meeting summary, including a meeting agenda, attendee location, remote-only meeting, audio/video quality and reliability, video usage, and meeting size. The model and survey results give a quantitative understanding of how and where to improve meeting effectiveness and inclusiveness and what the potential returns are.

Motivated by the email survey results, we implemented a post-meeting survey into a leading computer-mediated communication (CMC) system to directly measure meeting effectiveness and inclusiveness (during COVID-19). Using initial results based on internal flighting we created a similar model of effectiveness and inclusiveness, with many of the same findings as the email survey. This shows a method of measuring and understanding these metrics which are both practical and useful in a commercial CMC system. By improving meeting effectiveness, companies can save significant time and money. Improving meeting inclusiveness is hypothesized to improve meeting effectiveness, but also improves the working environment and employee retention at organizations.

CCS Concepts: • **Human centered computing** → Computer supported cooperative work; Empirical studies in collaborative and social computing

**ACM Reference format:**

Author's addresses: Ross Cutler, ross.cutler@microsoft.com; Yasaman Hosseinkashi, yahossei@microsoft.com; Jamie Pool, jamie.pool@microsoft.com; Senja Filipi, senja.filipi@microsoft.com; Robert Aichner, robert.aichner@microsoft.com; Yuan Tu, yuan.tu@microsoft.com; Johannes Gehrke, johannes@microsoft.com

# 1   INTRODUCTION

Computer-mediated communication (CMC) systems, especially VoIP and video conferencing systems, have transformed how organizations have meetings. Most recently such systems have enabled hundreds of millions of people to work from home during the COVID-19 pandemic, further increasing the need for such systems. A primary goal of these systems is to provide the most effective meetings possible for all participants. While there has been research in measuring and analyzing meeting effectiveness [1][2], this paper contains the first measurement and analysis of inclusiveness in meetings. This work provides evidence for the following research questions (RQ):

- RQ1: How much does inclusiveness impact meeting effectiveness?
- RQ2: What makes meetings more or less effective and inclusive?
- RQ3: How is the CMC system impacting meeting effectiveness and inclusiveness?
- RQ4: How can we practically measure meeting effectiveness and inclusiveness in a CMC system?

This paper first describes an email-based meeting survey study to measure meeting effectiveness and inclusiveness as well as multiple meeting attributes that can have a significant relationship with perceived effectiveness and inclusiveness in a CMC system. The goal is to understand how inclusiveness relates to effectiveness and how CMC system attributes can be used to predict and improve meeting effectiveness and inclusiveness.

The meeting attributes considered in this study were derived from several prior studies around meeting effectiveness and refined via a pilot study and validation analysis. These attributes are the presence of an agenda, pre-meeting reading materials, post-meeting summary, and meeting size. Remote communication attributes are using video, audio/video (AV) quality and reliability, meeting type (remote only or remote and a conference room), and join location (remotely or locally in the conference room). Section 3.1 contains details on how these attributes are suggested by previous studies to have a possible correlation with meeting effectiveness. Since our focus is to understand and improve meetings that are organized and run using CMC systems, we include AV quality and reliability and two separate questions to enable more granular analysis on remote communication facilities.

The focus of this study is twofold: 1) Examine the hypotheses about the perceived effectiveness relationship with meeting inclusiveness and other attributes listed above and 2) Find a mathematical model that reveals the structure of relationships suggested by the data. Note that the second task cannot be accomplished by a small sample. To the best of our knowledge, this is the first study that is using a sample as large as N=4,425 (N=3,290 after filtering) for this purpose, hence the first time that meeting effectiveness breakdown by multiple factors can be empirically evaluated.

Motivated by the results of the email survey and analysis, we implemented a post-meeting survey into a leading CMC system. The survey includes the primary response variables we want to measure (effectiveness and inclusiveness), as well as some qualitative questions to improve the meeting effectiveness and inclusiveness. We used the CMC system's internal telemetry data such as media used (audio, video, screen sharing), meeting size, predicted AV quality, and reliability to create a model of meeting effectiveness and inclusiveness. We show that the preliminary results are similar to results from the email study.

The CMC system used in this study supports audio and video conferencing on major computing platforms (Windows, macOS, iOS, Android, and web browsers), and includes dedicated conference room systems device support. It supports high-definition video (1080P) and super-wideband audio for high-quality AV experiences, and screen sharing for document or application sharing. The system supports 1-to-1 and group meetings, and ad-hoc and scheduled meetings.

This paper makes the following contributions:

1. We provide a quantitative model of the relationships of meeting effectiveness, inclusiveness, comfortableness to contribute, and participation based on these factors: pre-meeting communication sent, post-meeting summary sent, meeting agenda included, attendee location, remote-only meeting, AV quality, reliability, video usage, and meeting size. To the best of our knowledge, this is the largest survey study we are aware of (by a factor of four). It is also the first study that includes meeting inclusiveness, AV quality, reliability, remote/local participants, comfortableness contributing in the meeting, and pre-meeting reading (pre-reads).
2. We show that measuring meeting inclusiveness is possible and practical via surveys. Such surveys can thus be conducted periodically by organizations, or integrated into the CMC system and automatically delivered after meetings.

The meeting effectiveness and inclusiveness model and survey results show which areas of the CMC system need to be improved to increase meeting effectiveness and inclusiveness, which are discussed in Section 5. Improving meeting effectiveness can directly save organizations both significant time and money [24]. Improving meeting inclusiveness is hypothesized to improve meeting effectiveness (we show that inclusiveness and effectiveness are correlated, but don't show causality), as well as improve the working environment and worker retention in organizations [31].

## 2   RELATED WORK

Meetings have been extensively studied in the academic literature (a recent review is given in [24]). One method to understand what drives meeting effectiveness is to first create a survey for meeting participants that include parameters considered to be related to effectiveness, and then to analyze the survey to see if the hypotheses are correct. This process is called a *meeting design characteristics study*. A meeting design characteristics study by Leach et al. [21] shows that the agenda, quality of facilities, and ending on time were found to be correlated to perceived effectiveness (N=958 survey). Attendee involvement served as a key mediator variable in the observed relationships. Another meeting design characteristics study with even more factors is given by Cohen et al. [8] for meeting quality (N=367 survey). This study shows the top significant factors for meeting effectiveness are meeting space size, starting promptness, lighting quality, and organization type. Allen et al. [2] show that meeting satisfaction predicted employee empowerment, and information availability partially mediated this effect. Geimer et al. [13] asked meeting attendees how to improve meeting effectiveness. The results suggest that employees are often invited to meetings of little relevance and many organizers fail to apply fundamental meeting design practices.

Agendas have been shown to be correlated to meeting quality [8] and meeting effectiveness [7,21]. Similarly, post-meeting summaries (minutes) have been shown to be correlated to meeting quality [8] and meeting effectiveness [21]. Allen et al. [1] links pre-meeting discussions with meeting effectiveness.

Meeting size has been shown in [8] and [3] to be negatively correlated to meeting effectiveness. Odermatt et al. [27] showed uncivil meeting behaviors to be negatively correlated to meeting satisfaction and effectiveness.

While there has been significant research in better understanding meeting effectiveness, there is no common consensus on how to measure meeting effectiveness. For example, Garcia et al. [7] measure meeting effectiveness by the percent of agenda tasks that are completed. Standaert et al. [37] provides 19 business meeting objectives and used a survey with a 5-point scale (1: Not at all effective to 5: Very effective) on how different meeting modalities achieved the business meeting objectives for each of the 19 business meeting objectives. Nixon et al. [26] measure meeting effectiveness by two items, goal attainment and decision satisfaction. Rogelberg et al. [33] measure meeting effectiveness for meetings in a typical week using a 6 question survey (5-point scale) which include: "achieving your own work goals", "achieving colleagues' work goals", "achieving your department–section–unit's goals", "providing you with an opportunity to acquire useful information", "providing you with an opportunity to meet, socialize, or network with people" and "promoting commitment to what was said and done in the meeting." Leach et al. [21] measure meeting effectiveness for meetings in a typical week with a 3 question survey (5-point scale): "achieving your own work Goals", "achieving your colleagues' goals" and "achieving your department's / section's / unit's goals."

The number of studies that show significant value in video conferencing over audio conferencing is remarkably sparse and mixed. Veinott et al. [40] showed that video helps non-native speakers better negotiate than audio-only conferencing. However, Habash [15] showed that video added little or no additional benefit over audio-only conferencing for group perception and satisfaction in distributed meetings. Instead of measuring a task metric or satisfaction, Daly-Jones et al. [9] showed that video does improve conversational fluency and interpersonal awareness over audio-only meetings.

Sellen [34] showed when remote participants join an audio or video conference with a conference room, the conference room participants produced more interruptions and fewer formal handovers of the floor than the remote participants. Video did not improve the interruption or handover rate for remote participants compared to audio-only. Standaert et al. [37] showed that telepresence systems improved meeting effectiveness over audio and video conferencing systems. Tang et al. [38] showed that usage of a video conferencing system drops significantly when the video feature was removed, and analysis showed the video was used to help mediate their interaction and convey visual communication.

While there are many studies showing gender bias in speaking and interruption rates [11,18,22], both [8,21] show that gender is not correlated to meeting effectiveness. Triana et al. [39] showed women felt more included when [20][27]CMC was used before face-to-face meetings compared to doing face-to-face meetings and then CMC meetings; the order of communication media influenced the perceived inclusiveness, which influenced their participation. Guo et al. [14] showed that traditional face-to-face meetings outperformed videoconferencing teams when both teams had the same team-building experience. However, a dialogue-based framework can be used to help virtual teams to perform as effectively as traditional face-to-face teams.

Davidson [25] reviews studies on how trust and member inclusion are communication factors to foster collaboration in teams, though not meetings specifically. There are many guides on how to have inclusive meetings, e.g., [30], though remarkably no research that we are aware of that actually measure inclusiveness.

We are not aware of previous studies on measuring meeting inclusiveness, but inclusiveness has been extensively studied for organizations. Ashikali et al. [5] used a large scale (N=10,976) employee survey to build a structural equation model that shows how transformation leadership and diversity management correlate to an inclusive culture of an organization. A six-question survey is used to measure the inclusive culture of an organization. In [12] (Chapter 1) Ferdman defines a similar set of six experiences of inclusion for organizations. Pearce and Randel [29] defined a three question survey on measuring team inclusion. Rice et al. [32] conducted studies to further show how organizational inclusiveness relates to supervisory inclusiveness, and how that relates to citizenship behavior and affective commitment. In [12] (Chapter 17) Lukensmeyer et al. provide a list of important characteristics of a meeting that is truly inclusive which we discuss in more detail in Section 3.1.2.

To our knowledge, there are no previous studies of meeting effectiveness that include meeting inclusiveness, AV quality and reliability, remote/local participants, comfortableness in participating, or pre-meeting written communication.

In addition to the structural aspects of meetings, which we address in this paper, there has been significant work done on remote collaboration and the non-structural aspects of collaboration. Olson [28] provides a summary of recommendations for effective remote collaboration and best practices for remote meetings. Woolley [6,42] studied the collective intelligence of groups and showed that groups with more equal distribution of turn-taking and groups with more equal distributions of gender had a higher collective intelligence. Lykourentzou [23] studied how personalities affect crowd-sourced teams and found that teams without a surplus of leader-type personalities exhibited less conflict and their members reported higher levels of satisfaction and acceptance. Kulkarni [20] study massive online classes and found that the more geographically diverse the discussion groups, the better the performance of the students. Kiesler and Sproull [36] studied electronic mail systems and showed how it increased the flow of information in organizations, and in particular reducing social contexts such as location, distance, time, organizational hierarchy, age, and gender. Further discussion of these non-structural aspects is given in Section 5.

## 3  EMAIL SURVEY

### 3.1  Survey Design

The goal of the email survey was to measure two response variables, meeting effectiveness and inclusiveness, and the design characteristics that we thought were correlated to them. We wanted to keep the survey relatively short to increase the survey response rate. In particular, to measure the response variables we used direct questions rather than several indirect questions that were related to effectiveness and inclusiveness. The most related surveys to this study were done by Leach et al. [21] and Cohen et al. [8]. A comparison of their survey measurements with our survey is given in Table 1. Our complete email survey is given in Section 6.

Per our company's survey guidelines, we did not include demographics in the survey such as gender, age, race, and income. Some of these demographics may be correlated to effectiveness and inclusiveness and should be added in future surveys to make the model developed in Section 3.3.5 more complete.

In order to keep the survey relatively short we did not include the following measures by Leach [21] or Cohen [8] shown to be statistically significant: Full or part time, organization type, lighting quality, meeting space size, verbal agenda, end on time, and start on time.

Including all of these would have made the model developed in Section 3.3.5 more complete, but we would have received less data to validate it due to having a longer survey.

Due to the COVID-19 pandemic, the number of online-only meetings has dramatically increased. Many of the features in Table 1 will change in importance during this move to online systems. In Table 2 we give our expectations of these feature changes which should be validated in a future study.

In a pilot study, a 17-question survey with free-text answers for questions 15, 16, and 17 were designed and ran. This resulted in 800 responses. The answers to the free text questions were processed and the top suggestions were used for these questions, with a free-text option to handle other possible answers. The result of the pilot study is the 17-question survey available in Section 6.

*3.1.1 Measuring meeting effectiveness*

For this study, we measure meeting effectiveness with survey question #3 (For this meeting: How effective was the meeting at achieving the business goals?). This does not include decision satisfaction or a breakdown of whose goals were achieved. It represents the least common denominator of the above methods of measuring meeting effectiveness: meeting goal attainment.

*3.1.2 Measuring meeting inclusiveness*

In [12] Chapter 17 Lukensmeyer et al. provide a list of important characteristics of inclusive meetings (IM), which are related to characteristics of inclusive organizations:

- IM1: People of different backgrounds and points of view are treated well and feel comfortable.
- IM2: Participants know what to expect and how their input will contribute to the decision.
- IM3: Everyone has an opportunity to be fully engaged in all aspects of the discussion and deliberation.
- IM4: Each person's views are fairly recorded and taken into account.

Our survey question #5 (How comfortable did you feel contributing to the meeting?) and #15 (f. Including all required participants) partially cover IM1, though it does not include asking how diverse the attendees are or if they were treated well. For IM2 survey questions #10 (Did the meeting have an agenda with the meeting purpose and goals in the meeting invitation?) and #11 (Did the meeting have any pre-meeting reading sent out before the meeting (e.g., slides, documents)?) should cover what to expect, though a question could be added on how their input would be used to achieve the meeting goals in #10. For IM3 survey question #9 (How much did you participate in the meeting?) measures how much each participant actually participated, which we feel is more meaningful than the opportunity to participate. We did not have a survey question that addressed IM4. Future work would be appropriate to study how much it improves on the completeness of the effectiveness model in Section 3.3.5.

In the pilot study, 1.0% of respondents did not understand what "inclusive" meant for question #4 (How inclusive was the meeting?). Therefore, in the larger study, we included the definition "In an inclusive meeting everyone gets a chance to contribute and all voices have equal weight."[1]. As a result, we did not receive any questions asking what an inclusive meeting is, and the ratings for Question 4, option c ("Neither inclusive nor exclusive") dropped from 14% to 10% (significant with $p < 0.01$), suggesting that respondents better understood the question. Option c presents a neutral answer which is selected mostly when the question is perceived as not strictly relevant to experience. The inclusiveness question, as defined above, should be relevant to any group meeting if understood correctly. Therefore, we interpret the drop in option c from 14% to 10% as a sign that respondents had more clear understanding about the question.

---

[1] Found in https://www.atlassian.com/blog/teamwork/how-to-run-inclusive-meetings.

Table 1: Comparison of survey measurements included in Leach [1] and Cohen [2]. * is significant with p < 0.05, ** with p < 0.01 with meeting effectiveness.

| Design characteristic | Feature | Leach | Cohen | Ours |
| --- | --- | --- | --- | --- |
| **Attendee** | Age | N | Y | N |
| | Comfortable contributing | N | N | Y** |
| | Country | Y | N | N |
| | Education level | N | Y* | N |
| | Full, part time | Y** | Y | N |
| | Gender | Y | Y | N |
| | Hours worked | N | N | N |
| | Inclusiveness | N | N | Y** |
| | Involvement | Y** | N | Y** |
| | Job level | Y | Y* | N |
| | Meeting size | Y | Y* | Y |
| | Organization size | Y | N | N |
| | Organization type | N | Y** | N |
| | Supervisory status | N | Y | N |
| | Time employed | N | Y | N |
| **Physical** | Audio/video quality | N | N | Y** |
| | Audio/video reliability | N | N | Y** |
| | Facilities (overall) | Y | N | N |
| | Join location | N | N | Y** |
| | Lighting quality | N | Y** | N |
| | Meeting space size | N | Y** | N |
| | Meeting type (remote only, etc) | N | N | Y** |
| | Noise level | N | Y | N |
| | Refreshments | N | Y* | N |
| | Seating arrangement | N | Y | N |
| | Temperature comfort | N | Y* | N |
| | Video | N | Y | Y |
| **Procedural** | Agenda: verbal at meeting | Y** | N | N |
| | Agenda: written in advance | Y** | Y* | Y** |
| | Agreement use (rules) | N | Y | N |
| | Chairperson | Y | Y* | N |
| | Meeting type (content) | Y | N | N |
| | Minutes | Y** | Y | Y |
| | Pre-meeting communication | N | N | Y* |
| | Recording of the meeting | N | Y | N |
| **Temporal** | Break use | N | Y | N |
| | End on time | Y** | Y* | N |
| | Meeting duration | Y | Y | N |
| | Start on time | Y | Y** | N |

Table 2: Expected changes to feature importance for online only meetings

| Design characteristic | Feature | Change to online only |
|---|---|---|
| **Attendee** | Job level | Less correlated as online meetings can lower the boundaries set by different job levels |
| **Physical** | Lighting quality | More correlated as home lighting quality will have higher variance than conference rooms |
| | Refreshments | Less correlated since all home users have refreshments |
| | Temperature comfort | Less correlated since home users can set their room temperature for themselves |
| **Temporal** | Start on time | Less correlated since online meetings have no travel time |
| | End on time | Less correlated since online meetings have no travel time |

### 3.2 Executing the survey

To conduct the main study, the survey was sent to 16,000 employees selected randomly but excluding executives and their direct reports. A web-based survey tool was used to conduct the survey anonymously. A total of N=4,425 responses were received in a one-week period. Only responses where the user had a meeting within the last week were used in the analysis. This threshold was applied since we observed a positive correlation between the time-since-last survey and average ratings during the pilot: the longer the survey time from the meeting, the more likely to give a high rating.

In addition to the above, we also removed responses when the meeting size is larger than 50. This is to exclude special types of meetings that are not designed for collaboration (e.g., all-hands meetings) and hence not of the focus of this work. For similar reasons, we dropped the survey response if meeting type = 'other' or if the online collaboration system is not used to schedule or run the meeting. Applying all these filters and removing incomplete surveys leaves N=3,290 for analysis.

### 3.3 Analysis of Survey Results

The main goal of this study is to understand the relationships of effectiveness and inclusiveness with multiple attributes as described in Section 3.1. After addressing data preparation and validation steps, we begin with descriptive statistics and exploratory analysis to understand the data and initial measures of correlation in Section 3.3.3. Then we present the methodology that we developed for modeling effectiveness and inclusiveness based on exploratory analysis results. Sections 3.3.5 and 3.3.6 are dedicated to this topic.

*3.3.1 Survey Encoding*

Many of the survey questions are set up with a five-point scale rating, e.g., 5: Very Effective, 4. Effective, 3: Neither effective nor ineffective, 2: Ineffective, 1: Very ineffective. There are two different aggregations possible for summarizing the answers to these types of questions:

- Average rating: treating the responses as integer values between 1 and 5 and taking the linear average.
- Top two score proportion: count of 4 and 5 ratings divided by the total number of ratings.

Most of our analysis is focused on using the second aggregation. Hence the metric "Effective" for example, is the proportion of ratings that are either "Effective" or "Very effective". Similarly, Inclusive is the proportion of 'Inclusive' and 'Very inclusive'.

Question 9 'How much did you participate…' has a four-point scale survey answer. We define the metric *Participation* as speaking more than once in the meeting, i.e. 'I spoke up many times' or 'I spoke a few times.'.

For dichotomous scale questions such as the usage of video and presence of an agenda, one variable is generated which holds the value 1 for 'Yes' and 0 for 'No'. The rest of the questions are encoded using each response as a binary variable. For example, for question 2, we generated three different variables for meeting types: remote only, conference room only, remote, and conference room. Values with small samples are exempted. For questions where multiple encoded variables were generated, such as meeting type, we only show the ones that came out as significant in the final model. For example, the meeting type encoding that is presented in the results tables and graph is *RemoteOnly*, a binary variable to distinguish whether a meeting room was involved in the meeting or not.

The only integer variable in this analysis is the meeting size. Figure 1 shows the empirical distribution of the meeting size for this data. The most popular meeting size (the mode of the distribution) is 6 participants, while there is a long tail toward larger meetings. The 25$^{th}$ percentile of the meeting size is 5. We use this value as a criterion to distinguish large vs small meetings: meetings with 5 or less participants are considered small and meetings with 6 or more participants are considered as large. This way we add a dichotomous measure for meeting size: *MeetingSize5p*. Figure 1, shows the effectiveness score by meeting size buckets. Later, we will show how *MeetingSize5p* is related to effectiveness, inclusiveness, and participation.

Table 7 contains the encoding information about all variables that were used in the model.

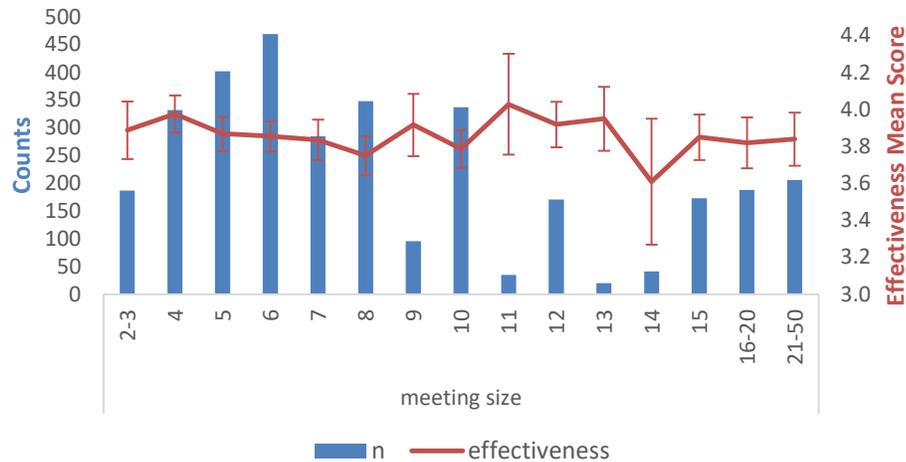

Figure 1: Effectiveness by meeting size

*3.3.2 Response Rate and Non-Response Bias Analysis*

The response rate for the main survey was 27%,. If these people are different from the employees who responded in terms of factors that matter to the survey, then the sample will not be representative of the target population. This bias is referred to as non-response bias and can be formally detected by comparing the distribution of known factors between respondents and the target population.

The target population in this study consists of all remote collaboration meetings at the company where we performed the survey. There are multiple variables available for the target population that we used for testing non-response bias. The CMC system telemetry data contains the number of devices in a meeting, in addition to the types of media each device sends and receives. The number of speakers in a meeting is approximated by the number of devices in the call where the microphone is muted for less than 95% of the meeting duration. These telemetry features are compared with the survey response questions for participation rate, video usage, and the number of participants. The resulting distributions are shown in Figure 2. A chi-squared test showed the telemetry and survey metrics differed with a $p < 0.01$. However, some difference in the distributions was expected since the definitions for the categories have a slight variation between the two populations. The overall difference is not large enough to require the resampling of the data (the delta < 15% for all metrics).

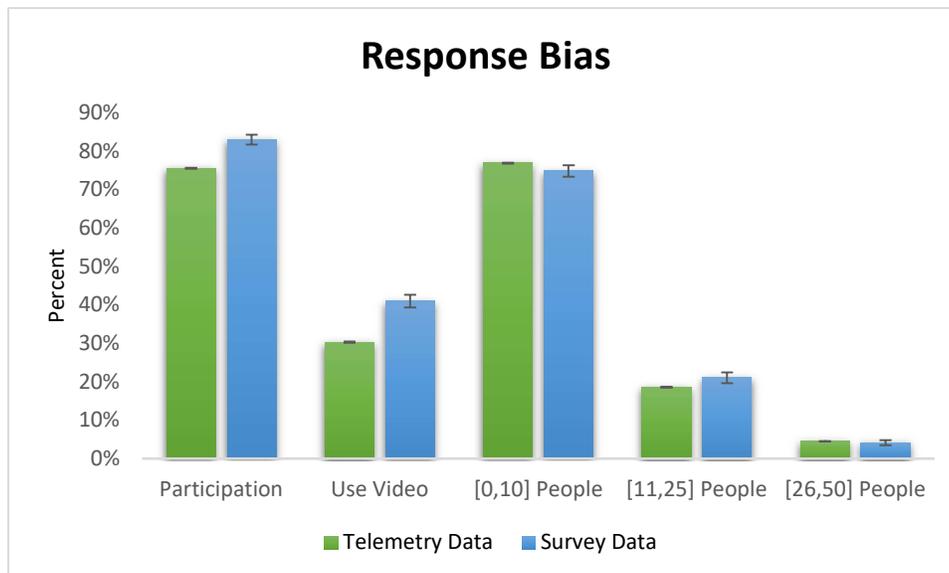

Figure 2: Bias check visualization (comparison of survey results to telemetry of all calls)

*3.3.3 Exploratory Analysis of Survey Results*

In this section, we provide descriptive statistics on the survey data. Table 3 shows the rate of each variable from the survey and its relationship with effectiveness. From the distribution of variables in row 'Rate' in Table 3, we see that more than 80% of meeting experiences were rated effective. 17% of meetings were perceived as ineffective which is very costly for the company just based on wasted employees' salaries. While 85% of raters expressed that they felt comfortable contributing to the meeting and 83% participated in conversation more than once, still 20% did not rate the meetings inclusive.

The 'Gain' rows show the absolute change in the probability of *Effective* and *Inclusive* by each factor if considered individually (using a single predictor model). The model is a simple logistic regression to predict *Effective* or *Inclusive*. For each meeting attribute, we fit a separate model. The 'Gain' value is marked if the model slope coefficient is statistically significant. As expected, *Inclusive* as well as *Participation* and *Comfortable* each show an individually significant relationship with *Effective*. The next level of significant relations is shown to be *Agenda* and *Postmeet* with both *Effective* and *Inclusive*. Our holistic model in Section 3.3.5 will help discover more information about *Video*. The single predictor model does not show any significant relation with the join type. This is measured by the variable *JoinRemotely* that is 1 if the respondent joined the meeting remotely and is 0 if the respondent joined locally in the conference room. For remote only meetings, this attribute is 1 for all participants. In the in-client survey that was conducted during the COVID-19 time period, all meetings are remote only and the variable *JoinRemotely* is dropped from the analysis. Finally, small meetings (less than 5 participants, *MeetingSize5p*) and remote only meetings (*RemoteOnly*) are shown to be more inclusive but not more effective.

Table 3: Survey summary statistics. Statistically significant at p < 0.05 marked by * and p < 0.01 marked by **

| | *Effective* | *Inclusive* | *Participation* | *Comfortable* | *MeetingSize5p* | *JoinRemotely* | *Video* | *Agenda* | *Premeet* | *Postmeet* | *Quality* | *Reliability* | *RemoteOnly* |
|---|---|---|---|---|---|---|---|---|---|---|---|---|---|
| **Rate** | 0.83 | 0.80 | 0.83 | 0.85 | 0.72 | 0.58 | 0.41 | 0.75 | 0.28 | 0.51 | NA[2] | NA | 0.30 |
| **Effective Gain** | | 0.04** | 0.01** | 0.04** | -0.01 | 0.00 | 0.01 | 0.02** | 0.02* | 0.02* | 0.01** | 0.01** | 0.01 |
| **Inclusive Gain** | | | 0.03** | 0.06** | -0.03** | 0.00 | 0.00 | 0.02** | 0.02 | 0.02** | 0.02** | 0.02** | 0.05** |

We also analyze the pairwise relationship between meeting attributes using a Chi-square test of independence, shown in Table 4. In this table, the strength of the correlation between two categorical attributes is measured by the odds ratio (OR) so that we can compare different pairs [10]. If *p* represents the probability of True or 1 for a binary variable, OR is defined as the rate $\frac{p}{1-p}$. OR is not as intuitive as probability itself but is a mathematically appropriate choice for comparing the strength of relationships between multiple pairs of binary variables. If there is a significant relation, the odds ratio is either higher than one (a positive correlation) or lower than 1 (a negative correlation).

We see a strong relation between *Quality* with *Reliability* and *RemoteOnly* with *JoinRemotely*. These are expected by the definition of these metrics, e.g., everyone must join remotely when the meeting does not involve a conference room. Also, the quality and reliability of audio and video are impacted by shared factors such as the CMC's network and device. Aside from these two pairs, the relationship between *Comfortable* with *Participation* is the highest. The next noticeable group in Table 4 is *MeetingSize5p*, *JoinRemotely* (join type), and *RemoteOnly* (meeting type: remote only or including a conference room).

---

[2] Not given as it is confidential CMC system statistic.

Table 4: Survey question pairwise relationship. Statistically significant at p < 0.01 marked by **

|  | Effective | Inclusive | Participation | Comfortable | MeetingSize5p | JoinRemotely | Video | Agenda | Premeet | Postmeet | Quality | Reliability |
|---|---|---|---|---|---|---|---|---|---|---|---|---|
| **Effective** | | | | | | | | | | | | |
| **Inclusive** | 3.46** | | | | | | | | | | | |
| **Participation** | 1.64** | 2.43** | | | | | | | | | | |
| **Comfortable** | 4.66** | 7.83** | 4.92** | | | | | | | | | |
| **MeetingSize5p** | 0.88 | 0.53** | 0.25** | 0.66** | | | | | | | | |
| **JoinRemotely** | 1.07 | 0.99 | 0.60** | 0.70** | 0.50** | | | | | | | |
| **Video** | 1.12 | 1.03 | 0.67** | 0.90 | 1.75** | 1.40** | | | | | | |
| **Agenda** | 1.54** | 1.6** | 1.04 | 1.52** | 1.61** | 0.79** | 1.34** | | | | | |
| **Premeet** | 1.29 | 1.22 | 1.30 | 1.22 | 1.24 | 0.92 | 1.20 | 3.21** | | | | |
| **Postmeet** | 1.28** | 1.30** | 1.40** | 1.44** | 1.28** | 1.10 | 1.15 | 2.34** | 2.31** | | | |
| **Quality** | 1.53** | 2.30** | 1.01 | 1.67** | 0.85 | 0.92 | 0.87 | 1.05 | 0.96 | 1.05 | | |
| **Reliability** | 1.50** | 2.23** | 0.92 | 1.72** | 1.05 | 1.00 | 0.93 | 0.95 | 0.90 | 1.08 | 26.14** | |
| **RemoteOnly** | 1.15 | 1.60** | 1.51** | 1.35** | 0.28** | 53.35** | 0.35** | 0.70** | 0.86 | 1.10 | 1.20 | 1.19 |

The multiple connections between variables in this table suggest that single predictor models (such as in Table 3) are not sufficient for understanding how these attributes impact meeting effectiveness and inclusiveness. Not only there are attributes such as *MeetingSize5p*, *RemoteOnly,* and *JoinRemotely* that need to be controlled when analyzing other predictors in the model, *Comfortable* and *Participation* stand out by high correlation between themselves and with several other attributes. In other words, the data is showing that *Participation* and *Comfortable* are naturally more like outcomes of other attributes rather than independent attributes themselves.

*3.3.4 Suggested improvements for effectiveness, inclusiveness, and comfortableness*

The results of questions 15-17 are given in Figure 3, Figure 4, and Figure 5. These results show some clear improvements for making meetings more effective, more inclusive, and more comfortable to participate. Some of these CMC system improvements are discussed in Section 5.

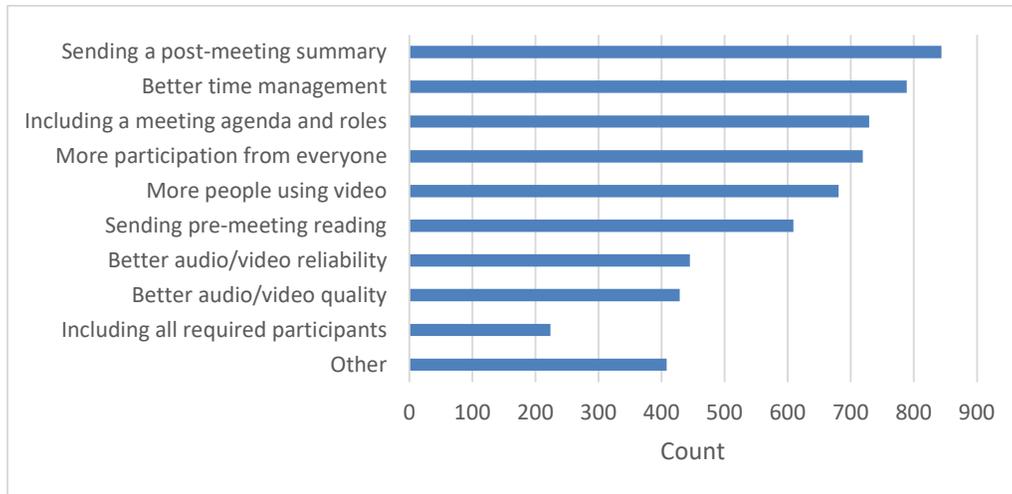

Figure 3: Responses to Question *What would have made the meeting more effective?*

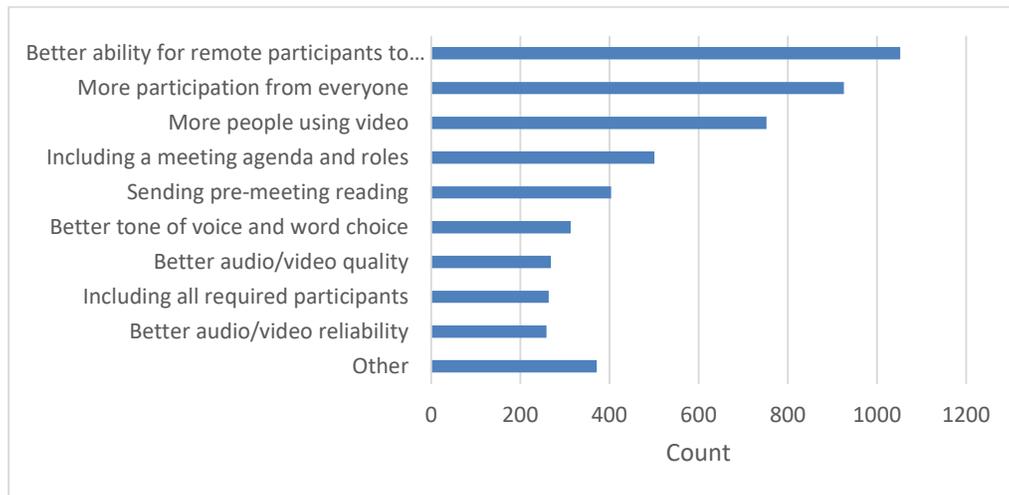

Figure 4: Responses to Question *What would have made the meeting more inclusive?*

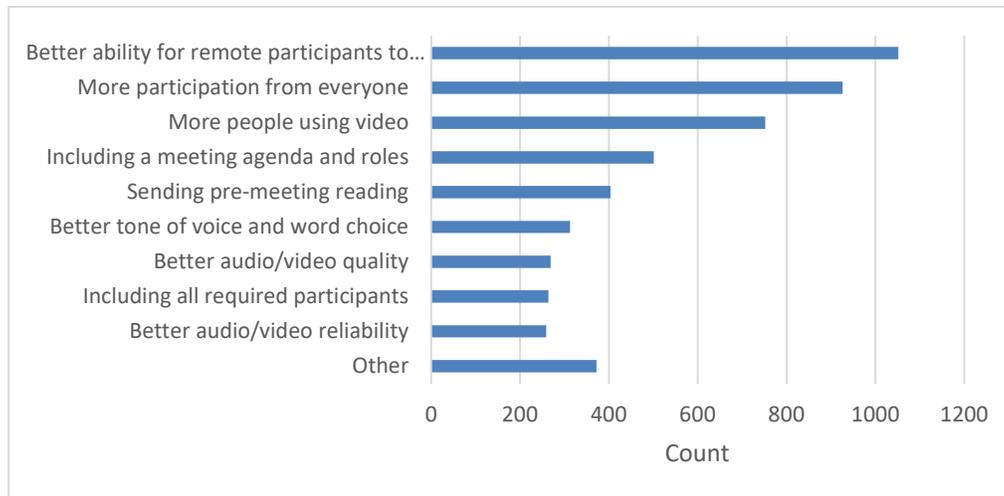

Figure 5: Responses to Question what would have made it more comfortable to participate in the meeting?

*3.3.5 Learning the graph of relationships to predict meeting effectiveness*

There are multiple options for incorporating all these variables into a single model. Strong predictive methods such as tree-based models are not aligned with the goal of this study due to the lack of interpretability. The exploratory analysis in the previous section gives strong motivation for a multi-layer or hierarchical structure for an interpretable model. This is also observed when trying to fit models that treat every attribute at the same level. For example, when predicting *Effective* using all meeting attributes via the Generalized Linear Model (GLM) [10], only two predictors are statistically significant: *Inclusive* and *Comfortable*. This is expected behavior given a strong correlation within the group of *Effective*, *Inclusive*, *Comfortable,* and *Participation*. As we move away from *Effective* and build models to predict the rest, more attributes turn significant as shown in Table 5. This observation reveals a specific relationship structure in the data that requires a more comprehensive modeling approach rather than a classical predictive model such as GLM or tree-based classifiers.

Table 5: Generalized linear model coefficients. Statistically significant at p < 0.01 marked by **

|  | GLM Coefficients | | | |
| --- | --- | --- | --- | --- |
|  | Effective | Inclusive | Participation | Comfortable |
| RemoteOnly | -0.04 | 0.39** | 0.42** | 0.45** |
| Inclusive | 0.77** |  |  |  |
| Participation | 0.00 | 0.29* |  | 1.51** |
| Comfortable | 1.21** | 1.89** | 1.51** |  |
| MeetingSize5p | 0.01 | -0.5** | -1.34** | -0.13 |
| JoinFromRemotely | 0.18 | -0.08 | -0.76** | -0.46** |
| Video | 0.11 | 0.21* | -0.15 | 0.14 |
| Agenda | 0.26* | 0.43** | -0.08 | 0.42** |
| Premeet | 0.14 | 0.09 | 0.23 | 0.04 |
| Postmeet | 0.08 | 0.07 | 0.31** | 0.23* |
| Quality | 0.15 | 0.49** | -0.08 | 0.25 |
| Reliability | 0.11 | 0.47** | -0.25 | 0.47** |
| AIC | 2747 | 2801 | 2554 | 2642 |

To this end, we apply graphical models as described in [16], Chapter 17. In a graphical model, nodes represent variables and edges carry information regarding the conditional probability distributions. These models are capable of learning a complicated network of relations. In practice, applying a graphical model involves two main steps: (1) learning the graph structure, and (2) estimating the parameters of the edges.

*3.3.5.1  Learning the graph structure*

Among the two steps, the former is more challenging for most cases. In general, optimizing a metric of the goodness of fit over the combinatorial space of graph structures is an NP-hard problem [16]. Wainwright et al. [41] give an algorithm based on $\ell 1$-regularized logistic regressions to estimate the graph structure. The algorithm involves finding each node's neighborhood using lasso regression and is shown to provide consistent estimations. Hoefling and Tibshirani [17] showed that this approach closely estimates the exact procedure through extensive simulations. The regularization also provides a robust selection of the graph structure since by forcing the smaller coefficients to zero, the selection of predictors for each outcome variable is expected to not be impacted by outliers.

In the current analysis, the problem of finding the graph structure is reduced to finding the optimum neighbors for each outcome variable. Following the approach in [41], we apply logistic regression with $\ell 1$- penalized log-likelihood optimization using the glmnet [43] package. If the coefficient estimated by the model for a particular outcome variable is non-zero, then there is a directed edge from that predictor to the outcome variable. Based on [43], glmnet optimizes for:

$$l(Y, BX) + \lambda[(1 - \alpha) \| B \|_2^2/2 + \alpha \| B \|_1] \qquad Eq.\ 1$$

where $B$, $X$ and $Y$ represent linear coefficients, input variables, and outcome variable respectively and $l$ stands for negative log-likelihood function. Parameter α balances between the lasso and ridge regression and is set to zero for our analysis. The parameter λ controls the strength of regularization: setting λ to zero results in no regularization, hence a dense graph with all edges present between predictors and outcomes. As we increase λ, the graph gets sparser. A few examples are shown in Figure 6.

The value of λ can be tuned locally using cross-validation to minimize misclassification error. We use 10-fold cross-validation to obtain local lambda values and pick a value that is close to the smallest local values. The reason is to minimize regularization to avoid missing out on weak correlations. By choosing $\lambda$, now a single graph structure is available. This concludes step (1) of the graphical model. The next step is to estimate the parameters that show the weight of each edge. These parameters are necessary to compare the strength of relationships and to predict the meeting effectiveness and inclusiveness.

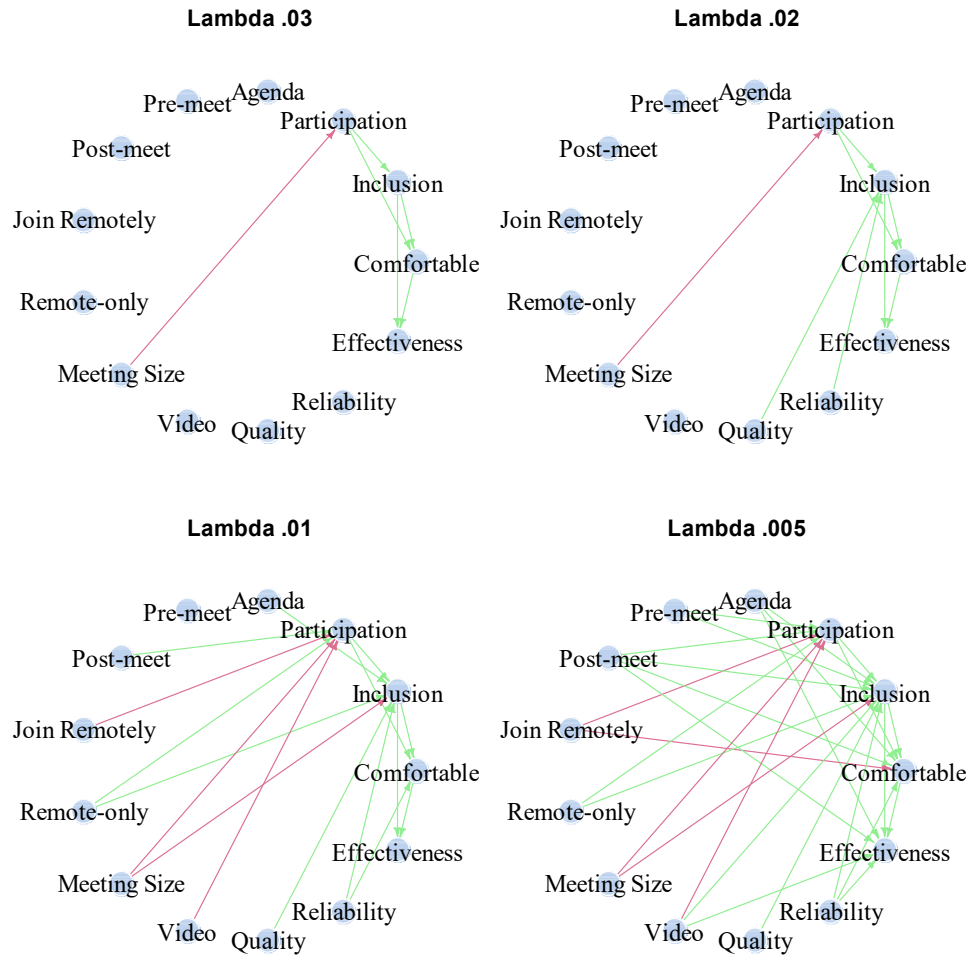

Figure 6: Meeting effectiveness model with 4 different versions of lambda (red is a negative correlation, green positive)

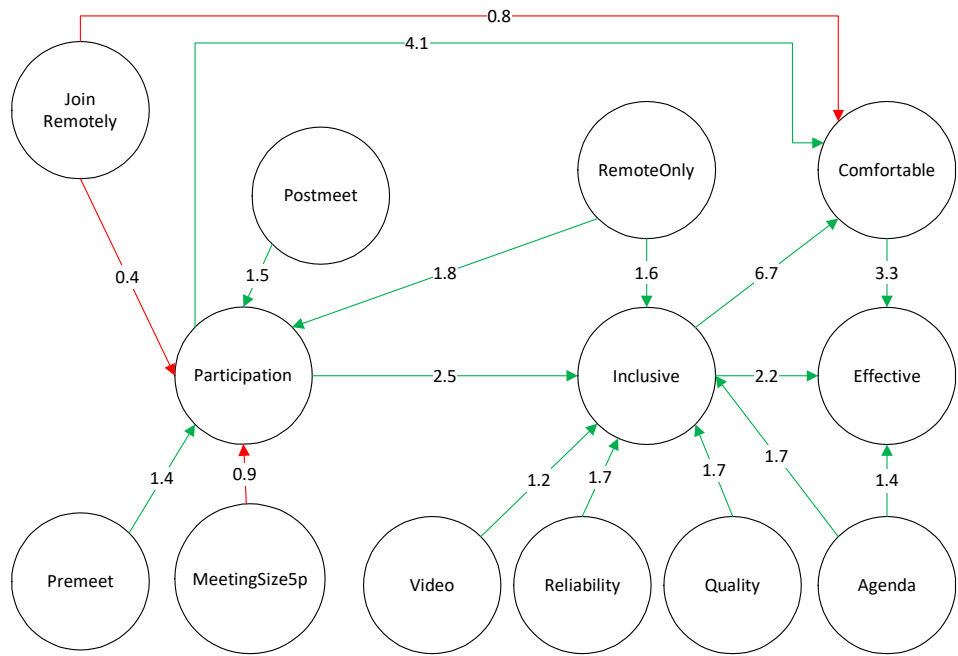

Figure 7: Multivariate model of effectiveness (red and green show negative and positive effects respectively)

*3.3.5.2 Estimating the parameters of graph structure*

Lasso coefficients from step (1) cannot be used for this purpose since there is no accurate way of estimating the uncertainty measures for them. Instead, we apply logistic regression without regularization. The linear coefficients of logistic regression ($\beta_i$) quantify the amount of increase in the log odds ratios (*log*OR) of the outcome as a result of the input variable being True *if* all other variables are kept constant. We use $\gamma_i = e^{\beta_i}$ as edge weights which represent the change in OR instead of *log*OR. We also use the p-values of the model coefficients $\beta_i$ to drop insignificant edges. This induces a secondary pruning on the fitted graph.

Figure 7 shows the fitted graph with 0.005 and all edges statistically significant at the 95% confidence level; the graph consists of four models. The detailed parameter values are presented in Table 6. Graph presented in Figure 7 consists of edges that are shown to be robust to sampling variations with one exception of Video – Inclusive edge. The detail about how to measure the consistency over sampling variation is discussed in Section 3.3.7. The reason that we exempted Video is that this feature is only relevant to a subset of data (video calls) and is sensitive to the bootstrap samples due to the variations of segment sizes for each sample.

Note that in logistic regression, the logit function (*log*OR) is modeled as a linear function of input variables. This means that the impact of one variable in the OR of the outcome can be measured without strong assumptions on other inputs. However, the probability of the outcome is not a linear function of inputs and does not have such nice properties as OR. For example, in the case of $logit(Effective) \sim \beta_0 + \beta_1 Inclusive + \beta_2 Comfortable + \beta_3 Agenda$, the impact of *Inclusive* on OR $\left(\frac{P(Effective=True)}{P(Effective=False)}\right)$ is given by $e^{\beta_1}$. This is saying that when *Agenda and Comfortable* are fixed, changing *Inclusive* from False to True will improve the odds of *Effective* by $e^{\beta_1}$ regardless of what value other inputs are fixed at. On the other hand, the impact on the probability of *Effective* is given by the change in

$$\frac{1}{1+e^{-BX}}$$ Eq. 2

where $BX = \beta_0 + \beta_1 Inclusive + \beta_2 Comfortable + \beta_3 Agenda$. As can be seen, the impact on probability is a function of two other variables. Therefore, the impact on probability is different given the status of other variables. While this can be a source of finer insight, it does not help with the overall comparison of impacts. For this goal, odds ratios are better mathematical tools. We offer both values in Table 6 where GLM Coefficients are $\beta_i$, Odds Ratios are the impact of each input in the odds of outcome ($e^{\beta_1}$), and Probability Impact shows the absolute change in the Output probability if the respective Input attribute is True, assuming that none of the other Input/Output attributes are True. For example, the first row shows that for a meeting that is not inclusive, not comfortable, and does not have an agenda, the probability of *Effective* can increases by 0.19 absolute percent if the meeting becomes inclusive. Note that the Probability Impact offers the expected change in outcome variable under specific circumstance which Odds Ratios measure the overall expected impact.

Table 6: Effectiveness Model Parameters

| Output | Input | Probability Impact | Odds Ratios | GLM Coefficient | Consistency (bootstrap) |
| --- | --- | --- | --- | --- | --- |
| *Effective* | *Inclusive* | 0.19 | 2.20 | 0.80 | 100% |
| *Effective* | *Comfortable* | 0.28 | 3.30 | 1.20 | 100% |
| *Effective* | *Agenda* | 0.08 | 1.40 | 0.30 | 89% |
| *Inclusive* | *Video* | 0.03 | 1.20 | 0.11 | 52% |
| *Inclusive* | *Agenda* | 0.13 | 1.70 | 0.51 | 100% |
| *Inclusive* | *Quality* | 0.13 | 1.70 | 0.52 | 100% |
| *Inclusive* | *Reliability* | 0.12 | 1.70 | 0.51 | 100% |
| *Inclusive* | *RemoteOnly* | 0.13 | 1.60 | 0.52 | 98% |
| *Comfortable* | *Inclusive* | 0.44 | 6.70 | 1.91 | 100% |
| *Comfortable* | *Participation* | 0.33 | 4.10 | 1.42 | 100% |
| *Comfortable* | *JoinRemotely* | -0.05 | 0.80 | -0.28 | 80% |
| *Comfortable* | *Agenda* | 0.05 | 1.30 | 0.25 | 76% |
| *Comfortable* | *Postmeet* | 0.04 | 1.20 | 0.19 | 64% |
| *Comfortable* | *Reliability* | 0.07 | 1.40 | 0.32 | 73% |
| *Participation* | *MeetingSize5p* | -0.13 | 0.90 | -1.40 | 100% |
| *Participation* | *JoinRemotely* | -0.07 | 0.40 | -0.92 | 100% |
| *Participation* | *Premeet* | 0.01 | 1.40 | 0.25 | 80% |
| *Participation* | *Postmeet* | 0.02 | 1.50 | 0.39 | 99% |
| *Participation* | *RemoteOnly* | 0.02 | 1.80 | 0.58 | 99% |

*3.3.6 Model Interpretation*

The model suggests that comfortableness and inclusiveness are the strongest drivers of meeting effectiveness. The odds of a meeting being effective is 2.2 times higher if it is inclusive and 3.3 times higher if attendees feel comfortable contributing (see RQ1). *Agenda* is only attributed outside of the group of outcome variables that stays directly correlated with *Effective* even when adjusted for *Inclusive* and *Comfortable*. The adjusted effect on the odds of *Effective* is relatively small (40% relative increase) but statistically significant after the final pruning of weak edges of the graph.

The model indicates that attributes such as AV quality and reliability, agenda, and using video are signs of an inclusive meeting (see RQ2). *Quality*, *Reliability*, and *Agenda* each can improve the odds of inclusiveness by 70% and increases its probability by the absolute value of 0.12 – 0.13 individually (see RQ3). Also, the odds of *Inclusive* is 60% higher if it does not involve a conference room! This confirms our observation in the exploratory step.

Note that *Postmeet* is correlated to both *Participation* and *Comfortable*. This relationship is not causal but rather indicates that meetings that include *Postmeet* communications are ones where attendees both feel more comfortable and able to participate. It may be that these meetings are well executed by a meeting organizer or facilitator who not only sent a summary of the meeting but also conducted the meeting better than average.

This model helps in understanding where we can invest in to improve meeting effectiveness and inclusiveness. For example:

- Meetings are most effective when:
    - attendees feel comfortable participating
    - they are inclusive
    - an agenda is included
- Meetings are most inclusive when:
    - an agenda is included
    - the AV quality and reliability is good
    - it is a remote-only meeting
    - attendees speak more than once
    - attendees see a video of others
- Attendees feel most comfortable contributing when:
    - the meeting is inclusive
    - they are not joining remotely
- Attendees participate most when:
    - pre-meeting reading is sent
    - they don't join remotely when the meeting is not remote-only or the meeting is remote only
    - the meetings are not large

Some of these relationships are not immediately obvious. For example, an agenda helps prepare meeting participants for the meeting and lets them know how they can participate (see IM2 in Section 3.1.2). Similarly, when pre-meeting reading is sent, participants can also prepare and therefore can participate more in the meeting discussion. The audio/video quality and reliability need to be good enough for participants to interact and participate; if the call quality is low or the call drops, participants will not be able to participate which is highly correlated to inclusiveness. In the in-client meeting survey (Section 4.2) reliability is correlated to both participation and inclusiveness. In larger meetings, participants will naturally participate less as the chance to talk during a meeting is 1/N where N is the number of meeting participants. Remote participants speak less than conference room participants since their audio/video experience is inferior to an in-room experience (especially the latency and reduced ability to read non-verbal communication), making it difficult to get the speaking floor. Finally, meetings that have post-meeting summaries tend to have more participation; we believe this is due to a selection bias in that meeting organizers that take the effort to send out post-meeting summaries also send agendas

*3.3.7 Model Completeness and Robustness*

We use this graph shown in Figure 7 as a descriptive model to provides insight into the drivers of effectiveness and inclusiveness. However, we are mindful that these insights are limited to the pool of attributes that we included in the model. We believe that a complete model for *Effective* and *Inclusive* will need more inputs such as organization type (e.g., sales, brainstorming, project review, etc.), meeting duration, the use of application sharing, and more behavioral signals about participants. One way of measuring how much this model is further away from a complete model is to measure its predictive power. A full model should be able to predict *Effective* and *Inclusive* ratings with high accuracy. We evaluate the model predictive power by the Receiver Operating Characteristic area under the curve (AUC). This metric is robust to sample imbalance and type I / type II error trade-off. *Effective* and *Inclusive* AUC computed in a Monte Carlo cross-validation (50 random splits) are 0.68 (+/- 0.05) and 0.74 (+/- 0.04) respectively. The sample size does not allow for a large enough test set, hence the standard deviation of the AUC values in the cross-validation process is relatively high. We do not have a proper benchmark to evaluate the AUC value for effectiveness and inclusiveness models, and while in general an AUC > 0.7 is considered good [35], there is definitely room for improvement. One method to improve the AUC is to include additional attributes in the survey. From Table 1 we are missing job level, meeting size, organization type, end on time, start on time, and meeting duration, which have all been shown in prior studies to be correlated to meeting effectiveness.

To evaluate the robustness of the model to sample variations, we use a bootstrap method: resample the data, infer and fit the graph and compare the edges with the graph that is fitted to the full data. The parameter $\lambda$ is fixed at $\lambda = 0.005$ for all iterations. The bootstrap sampling is repeated 500 times and we saved the final graph for each sample. Some of the edges that we observed in the full graph disappeared in some bootstrap graphs due to sample variation. Column 'Consistency' in Table 6 shows the rate of each edge showing up in the bootstrap graphs. The final model presented in Figure 7 contains only edges that have consistency higher than 80%. The only exception is Video which is discussed in detail below.

Most edges in Figure 7 are shown to be robust to sample variations, i.e. high bootstrap Consistency. The lowest consistency edge is *Video – Inclusive*. The impact of *Video* on inclusiveness seems to be overly sensitive to sampling variation. This suggests that the impact might be limited to a specific segment of the data and whenever that segment has a lower frequency, the effect becomes statistically insignificant. To find such segments, we used the available information about meeting type, join type, and meeting size, and whenever sample size allowed, estimated the impact of *Video* on *Inclusive* for segment combinations. Our search showed that the significant improvement in *Inclusive* by *Video* is most dominant in these scenarios: 1. Meetings with less than 5 participants (*Inclusive* OR is 1.8 times higher, p-value 0.01), 2. If the participant has joined locally to a meeting that has remote participants (*Inclusive* OR is 1.4 times higher, p-value 0.04). It should be mentioned that this deep dive analysis is possible due to the large sample size that has sufficient power for detecting statistically significant effects specific to the subpopulations.

## 4    IN-CLIENT SURVEY

The email survey gave us confidence that the perceived meeting effectiveness and inclusiveness can be measured and has meaningful relationships with attributes related to the CMC system used for conducting the meetings. As the next step, we have implemented a short end-of-meeting survey into the CMC system to collect Effectiveness and Inclusiveness ratings directly. These results are being used to complement our understanding of factors related to inclusiveness and effectiveness so we can prioritize features in the CMC system that help with improving them and detect potential obstacles imposed by this technology itself on these notions. Moreover, we can collect orders of magnitude more survey data with an in-client survey compared to an email survey, and the survey can be directly correlated to the CMC system telemetry to make the survey significantly shorter; specifically, we don't need to ask if video was used or was an agenda sent, but only the minimal response functions we are interested in.

As with any new feature in a product, this post-meeting survey is going through an internal testing and flighting phase. However, the initial results are promising and resemble the same patterns as in the email survey. This section briefly reviews the current results of this initiative.

**4.1 In-client survey design**

There were three main concepts strongly related to meeting effectiveness in our email survey analysis: *Comfortable*, *Inclusive*, *Participate*. Among them, participation as a binary indicator can be inferred from available CMC system telemetry without requiring a user rating or access to meeting content so it was excluded from the post-meeting survey. We also made a conscious choice to exclude *Comfortable*. The reason is to keep the survey as short as possible without loss of practically needed information. *Comfortable* is so strongly correlated with *Effective* and *Inclusive* that it plays almost as a duplicate of information and does not provide much practical benefit such as bringing new connections to the model.

The in-client survey is shown in Figure 8, Figure 9, and Figure 10, which are displayed to the CMC user after the meeting and in succession. The order of the problem areas in Figure 10 is randomized for each meeting to prevent an ordering bias.

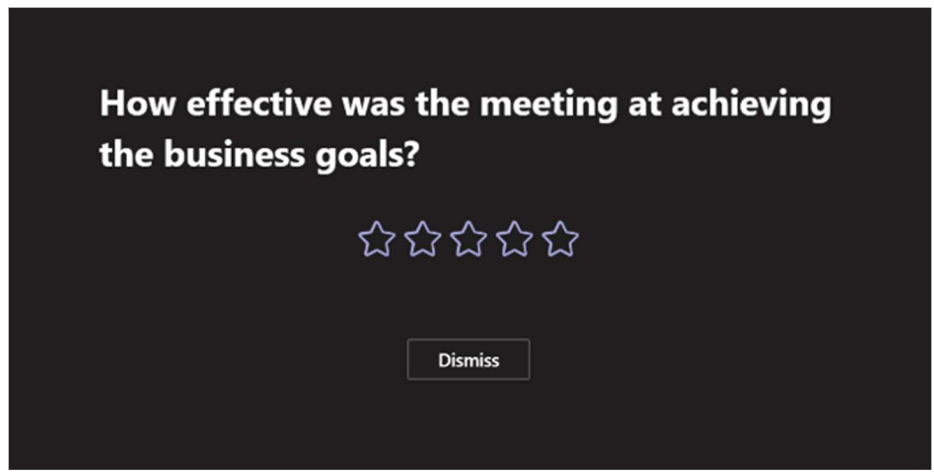

Figure 8: Post-meeting client survey #1

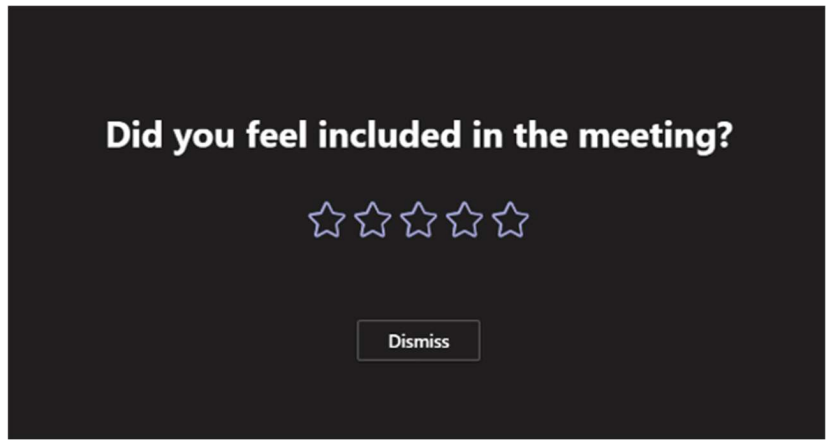

Figure 9: Post-meeting client survey #2

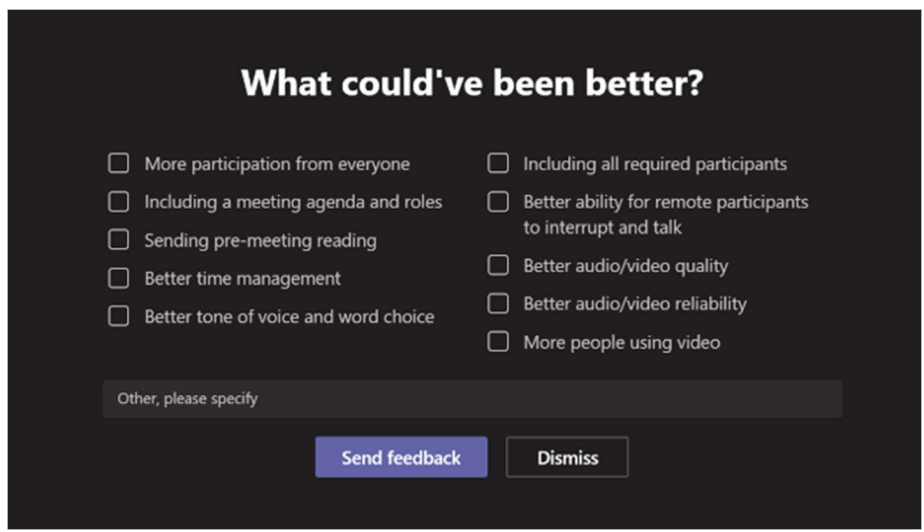

Figure 10: Post-meeting client survey #3

**4.2 In-client survey analysis**

We followed the same methodology described in Section 3.3.5 to develop the new model. Since this data is collected with respect to a specific meeting, a large pool of meeting attributes is available to be used for model development. The new attributes available for the model are: calls going through reconnect (*MidCallRel*), microphone failure (*MicFailure*), call drop (*Drop*), using application sharing (*ScreenShare*), and call duration. Call duration is broken down into three categories: less than 10 minutes (*duration0_10*), 10-20 minutes (*duration10_20*), and 20 min or longer (*duration20_*). This breakdown is determined by generating finer bins in the duration and combining the bins that were similar in terms of their relationship with *Effective* and *Inclusive*. Finally, call quality (*Quality*) is estimated using a separate classifier that consumes over 40 CMC system technical telemetry statistics to track video freeze, network issues, AV sync issues, etc.

We ran the algorithm using all available telemetry and let the data to choose which metrics should be used for the model. The available set of features for building this model is different from the survey model. For example, no indicator is available to detect whether the meeting included an agenda. The new telemetry, on the other hand, provided more detailed features about the call modality: in addition to an indicator about Video/ScreenShare application, we can also compute the % of call duration with Video for each participant. While the resulting graph is not identical to the email survey graph, it follows remarkably similar patterns. For example, in the new model participation is not directly asked from users, but rather inferred from the technical telemetry. But, like the email survey, this proxy of participation measured via telemetry demands to be a connection node between *Inclusive* and other attributes, instead of an independent predictor.

The fitted model based on N=17,287 over four weeks during-COVID-19 ratings is shown in Figure 11. This survey has a ~15% response rate and rating distribution is left skewed with 94% of ratings being 4 or 5 stars. In the updated model we observed a larger spectrum of coefficients compared to the email survey. All edges are statistically significant, but some are showing significantly larger coefficients and OR values than others. We believe this high variation may be a construct of data imbalance and data source mismatch. For example, *Inclusive* and *Effective* are subjective metrics and are showing a large correlation, while *Participation* and other metrics are objective telemetries coming from a different source. This might require more tuning and denoising before mixing them into one model. Also, during-COVID-19 meetings are all remote which was shown in previous results (email survey) to be more inclusive than other meeting types. Like the previous model, new data indicates a negative impact of increasing the meeting size on participation and a positive impact of Video usage on Inclusiveness. It also shows a similarly strong impact of using either Video or ScreenShare on Effectiveness As with any new feature, we are improving the data collections and expect the graph to mature as we collect more data. Overall, we think the in-client survey and analysis demonstrate a practical method to measure and understand meeting effectiveness and inclusiveness in a CMC system (see RQ4).

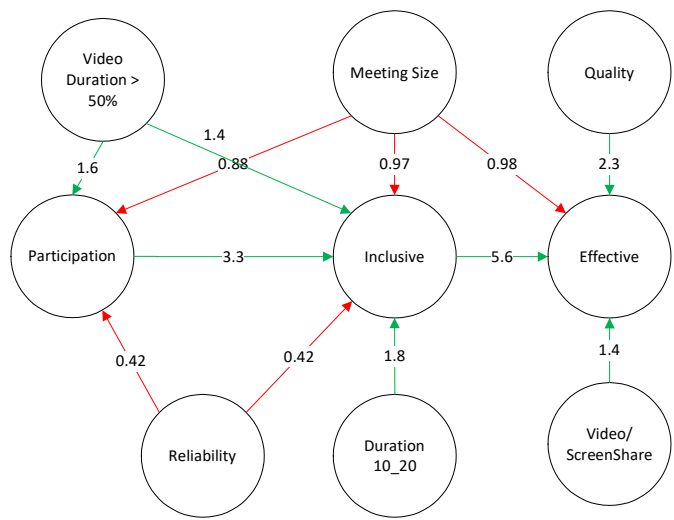

Figure 11: Effectiveness model from the in-client end of meeting survey

## 5  CONCLUSIONS

In this study, we conducted email and in-client surveys and created multivariate models of the relationships of meeting effectiveness, inclusiveness, comfortableness to contribute, and participation. The model and survey results can be used to answer the research questions given in Section 1.

The high-level results show that there is significant room for improvement for meetings to be more effective, inclusive, and where everyone feels comfortable participating. Some specific guidance for developers of CMC systems is given below:

- CMC systems should include surveys to measure meeting effectiveness and inclusiveness, which can be used to build models as shown in this paper to better understand how the CMC system relates to effectiveness and inclusiveness. In addition, the survey can be used in AB tests to measure improvements in the CMC system to improve these metrics.
- Video usage is correlated with more inclusive meetings. CMC systems should encourage video usage by making them the default modality, nudging users to use video, and reducing barriers to video usage such as reducing bandwidth, ensuring video doesn't degrade audio quality, and reducing fatigue due to poor quality video [4].
- Remote participation needs to be significantly improved, especially to help remote participants interrupt and get the speaking floor.
- CMC systems need tools to promote including agendas and premeeting readings in meeting invitations, as well as tools to create automatic meeting summaries.
- To improve larger meeting participation, CMC systems should not join auto-muted but create reliable/robust noise suppression technology so muting is not needed.
- Audio/video quality and reliability are critical for effective and inclusive meetings, so CMC systems need to continue improving it until it is shown that the quality and reliability no longer impact effectiveness and inclusiveness.

One confound to this study is the company surveyed may not be similar to many other companies or organizations. This can be addressed by running the survey across many different companies and generate one or more multivariate models. There may be different graph structures for different organizations, or possibly just a few general graphs that cover nearly all companies and organizations.

As a future area of research, we can use the survey for conducting AB tests to improve meeting effectiveness and inclusiveness by adding new features to our CMC system. The meeting effectiveness model allows for even more understanding of these metrics and will allow us to prioritize which parts of the CMC system to improve to increase meeting effectiveness and inclusiveness, and give feedback to the CMC system user on how to improve the effectiveness and inclusiveness of their own meetings or their organization's meetings.

In order to better predict meeting effectiveness and inclusiveness with our model, we plan to increase the input features in the in-client survey model so it is more complete, starting with missing significant features in Table 1. In addition, meeting participant attributes need to be included in our model, not just the structural attributes of the meeting. This includes personality attributions of the participants (e.g., is there a surplus of leader-type personalities in the meeting [23]), geographic location [20], as well as gender [6] and racial information. Based on [11,18,22] we expect to see gender bias in meeting participation rates. We are not aware of previous research on racial bias in meeting participation rates, but it would not be surprising to find that these also exist.

Additional research needs to be done to measure the survey biases such as non-response bias, survey exposure fatigue, telemetry loss bias, and the impact to the survey being anonymous or not. These are all important areas that are required for integrating the survey into a CMC system.

This study relies on a short survey that directly measures concepts such as inclusiveness and effectiveness. An alternative approach is to conduct the study with a larger survey with multiple questions targeting such concepts. Then inclusiveness and effectiveness and their relations can be inferred using Confirmatory Factor Analysis (CFA) and Structural Equation Models (SEM) [19]. This approach needs implementing a prior hypothesis about key contributors to effectiveness and inclusiveness and requires a larger survey which can cause a lower response rate and high non-response bias. It also is not applicable for the in-client survey that enables continuous monitoring of these metrics as key performance indicators of an organization due to the high cognitive load of filling out a large survey at the end of a meeting. However this approach can be used to help further validate the short direct survey method.

Improving meeting effectiveness and inclusiveness has significant financial benefits to organizations, as well as making their workplace a better environment to collaborate and retain employees. Our hope is this research improves the understanding of how to do this, and that future meetings get closer to the goal that all meetings achieve their business goals and that everyone feels included in their meetings.

## 6  APPENDIX

The survey questions used in the email survey are below:

1. When was the last meeting with 3 or more people that you participated in? (required)
    a. Today
    b. Yesterday
    c. 2-7 days ago
    d. More than a week ago
2. What type of meeting was it? (required)
    a. Group meeting in the conference room only
    b. Remote-only group meeting (no conference room)
    c. Distributed group meeting including conference room(s) and remote participants
    d. Other
3. For this meeting: How effective was the meeting at achieving the business goals? (required)
    a. Very ineffective
    b. Ineffective
    c. Neither effective nor ineffective
    d. Effective
    e. Very effective
4. How inclusive was the meeting? In an inclusive meeting everyone gets a chance to contribute and all voices have equal weight. (required)
    a. Not at all inclusive
    b. Not inclusive

   c. Neither inclusive nor exclusive
   d. Inclusive
   e. Very inclusive
5. How comfortable did you feel contributing to the meeting? (required)
   a. Very uncomfortable
   b. Uncomfortable
   c. Neither comfortable nor uncomfortable
   d. Comfortable
   e. Very comfortable
6. How many people attended the meeting (approximately)? (required)
7. Did you join in a conference room or remotely? (required)
   a. Conference room
   b. Remotely
8. Did you receive video of other participants in the meeting? (required)
   a. Yes
   b. No
9. How much did you participate in the meeting? (required)
   a. I only listened
   b. I spoke up once
   c. I spoke a few times
   d. I spoke up many times
10. Did the meeting have an agenda with the meeting purpose and goals in the meeting invitation?
    a. Yes
    b. No
11. Did the meeting have any pre-meeting reading sent out before the meeting (e.g., slides, documents)? (required)
    a. Yes
    b. No
12. Did the meeting have any post-meeting summary or action items sent out? (required)
    a. Yes
    b. No
13. If you used <anonymous audio/video communication application> in the meeting how was the audio/video call quality? (required)
    a. Excellent
    b. Good
    c. Fair
    d. Poor
    e. Very bad
14. If you used <anonymous audio/video communication application> in the meeting how was the call reliability (meeting join, call drops, screen share worked, etc.)? (required)
    a. Excellent
    b. Good
    c. Fair
    d. Poor
    e. Very bad
15. What would have made the meeting more effective? (not required, options shuffled)

      a. More participation from everyone
      b. Including a meeting agenda and roles
      c. Sending pre-meeting reading
      d. Sending a post-meeting summary
      e. Better time management
      f. Including all required participants
      g. Better audio/video quality
      h. Better audio/video reliability
      i. More people using video
      j. Other

16. What would have made the meeting more inclusive? (not required, options shuffled)
      a. More participation from everyone
      b. Including a meeting agenda and roles
      c. Sending pre-meeting reading
      d. Better tone of voice and word choice
      e. Including all required participants
      f. Better audio/video quality
      g. Better audio/video reliability
      h. Better ability for remote participants to interrupt and talk
      i. More people using video
      j. Other

17. What would have made it more comfortable to participate in the meeting? (not required, options shuffled)
      a. Including a meeting agenda and roles
      b. Sending pre-meeting reading
      c. Better tone of voice and word choice
      d. Better audio/video quality
      e. Better audio/video reliability
      f. Better ability for remote participants to interrupt and talk
      g. More people using video
      h. Other

Table 7: Email survey encoded variables used in the analysis

| Survey Question | Encoded Variable | Corresponding Survey Answer(s) |
|---|---|---|
| What type of meeting was it? | RemoteOnly | Group meeting in the conference room only |
| How effective was the meeting at achieving the business goals? | Effective | Effective, Very effective |
| How inclusive was the meeting? In an inclusive meeting everyone gets a chance to contribute and all voices have equal weight. | Inclusive | Inclusive, Very inclusive |
| How comfortable did you feel contributing to the meeting | Comfortable | Comfortable, Very comfortable |
| How many people attended the meeting (approximately) | MeetingSize5p | 0 if meeting size is less than or equal to 4, 1 otherwise |
| Did you join in a conference room or remotely? | JoinRemotely | Remotely |
| Did you receive video of other participants in the meeting | Video | Yes |
| How much did you participate in the meeting? | Participation | I spoke a few times, I spoke up many times |
| Did the meeting have an agenda with the meeting purpose and goals in the meeting invitation? | Agenda | Yes |
| Did the meeting have any pre-meeting reading sent out before the meeting (e.g., slides, documents)? | Premeet | Yes |
| Did the meeting have any post-meeting summary or action items sent out? | Postmeet | Yes |
| If you used <anonymous audio/video communication application> in the meeting how was the audio/video call quality? | Quality | Excellent, Good |
| If you used <anonymous audio/video communication application> in the meeting how was the call reliability (meeting join, call drops, screen share worked, etc.)? | Reliability | Excellent, Good |